\documentclass{iopart}
\usepackage{epsfig}
\usepackage{latexsym}

\begin{document}

\newcommand{\tbox}[1]{\mbox{\tiny #1}}
\newcommand{\half}{\mbox{\small $\frac{1}{2}$}}
\newcommand{\mbf}[1]{{\mathbf #1}}

\newcommand{\cn}[1]{\begin{center} #1 \end{center}} 
\newcommand{\hide}[1]{#1} 
\newcommand{\mboxs}[1]{\mbox{\small #1}} 
\newcommand{\mpg}[2]{\begin{minipage}[t]{#1cm}{#2}\end{minipage}}	
\newcommand{\mpb}[2]{\begin{minipage}[b]{#1cm}{#2}\end{minipage}}
\newcommand{\mpc}[2]{\begin{minipage}[c]{#1cm}{#2}\end{minipage}}


\title{Overview: Energy Absorption by Driven Mesoscopic Systems} 

\author{Doron Cohen}

\address{Department of Physics, Harvard University, Cambridge, MA 02138, USA}

\begin{abstract}
There are three regimes in the theory of energy absorption: 
The adiabatic regime, the linear-response (Kubo) regime, 
and the non-perturbative regime. The mesoscopic Drude formula 
for electrical conductance, and the wall formula for friction,  
can be regarded as special cases of the general formulation 
of the dissipation problem.   
The overview is based on a research report for 1998-2000.
\end{abstract}

The {\em wall formula} for the calculation of friction in 
nuclear physics~[1], and the {\em Drude formula} for the calculation 
of conductance in mesoscopic physics, are just two special 
results of a much more general formulation of `dissipation theory'. 
The general formulation is as follows: 
Assume a time-dependent chaotic Hamiltonian 
${\cal H}(Q,P;x(t))$ with $x(t){=}Vt$.
Assume also that $V$ is slow in a classical sense.  
For $V=0$ the energy is constant of the motion. For non-zero $V$
the energy distribution evolves, and the {\em average} energy 
increases with time. This effect is known as {\em dissipation}. \\

Ohmic dissipation means $d \langle {\cal H} \rangle /dt = \mu V^2$, 
where $\mu$ is defined as the dissipation coefficient.  
Ohmic dissipation, with an associated (generalized) Fluctuation-Dissipation 
relation for the calculation of $\mu$, can be established 
within the framework of classical mechanics~[2,3] using 
general classical considerations~[4].   
See [P2] for detailed presentation, and for discussion of 
validity conditions.  \\

\ \\
\begin{center}
\leavevmode
\epsfysize=2.0in 
\epsffile{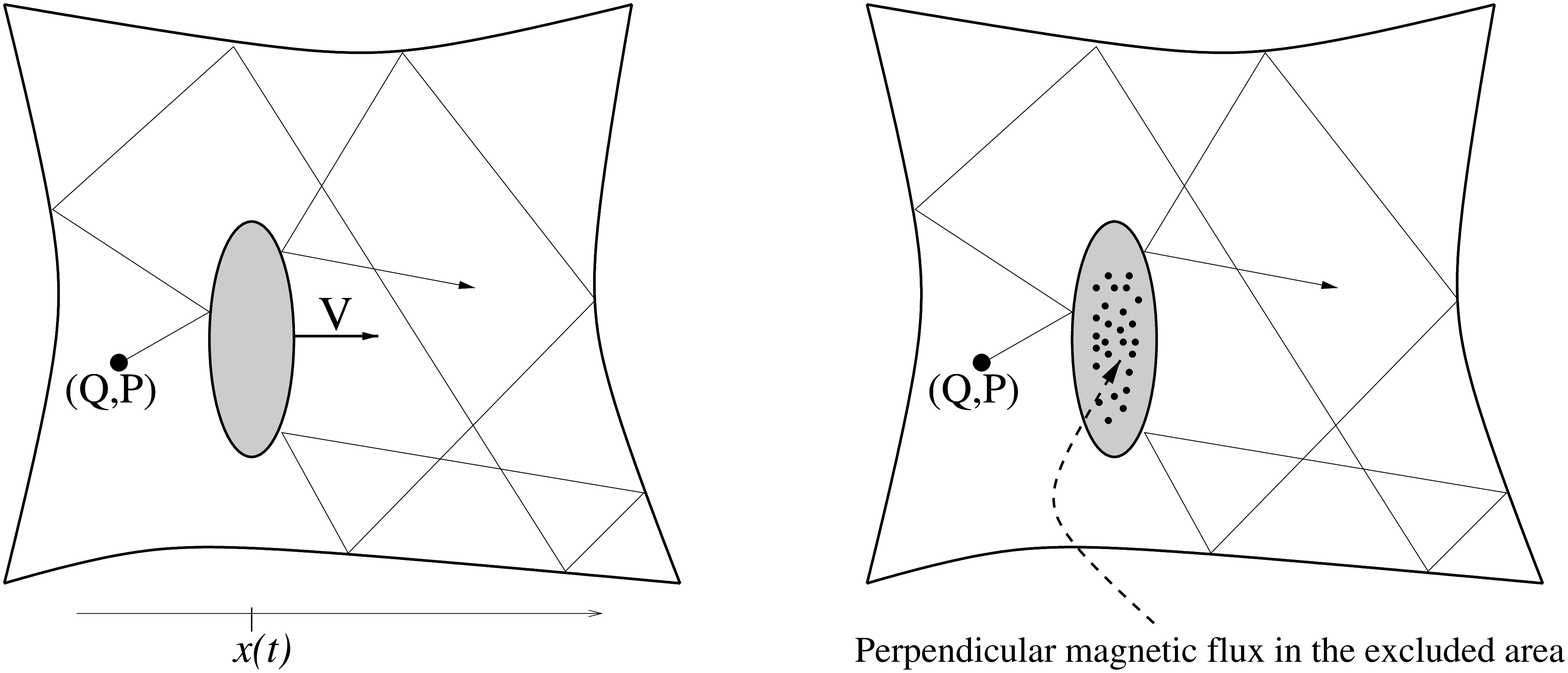}
\end{center}
{\footnotesize {\bf Fig.1.} Two examples where the general 
theory of dissipation can be applied. The left illustration 
is for the `wall formula', and the right illustration is for 
the `Drude formula'. See text for explanations.}  \\ 
  

The two leading examples where the above general formulation
can be applied are illustrated in Fig.1.
In case of the wall formula, $(Q,P)$ is a particle moving 
inside a chaotic `cavity', and $x$ controls the deformation 
of the boundary.  Ohmic dissipation (in the sense defined above) 
implies a friction force which is proportional to the velocity, 
where $\mu$ is the `friction coefficient', and $\mu V^2$ is 
the `heating' rate. 
In case of the mesoscopic Drude formula, $(Q,P)$ is a charged 
particle moving inside a chaotic `ring', and $x$ is the magnetic 
flux through the hole in the ring. 
Ohmic dissipation implies Ohm law, where $V\equiv\dot{x}$ 
is the electro-motive-force, $\mu$ is the conductance, 
and $\mu V^2$ is the `heating' rate. \\

The rate of energy absorption of driven chaotic cavities 
has been studied [P5,P6] with {\em A. Barnett} and {\em E. Heller}. 
For various reasons a satisfactory 
theory for the frequency dependent dissipation coefficient $\mu(\omega)$ 
has not been introduced in past studies of nuclear friction. 
It is somewhat surprising that to `shake' a cavity is not 
an effective way to heat up the `gas' inside it (Fig.2). \\

\ \\
\begin{center}
\leavevmode
\epsfysize=1.6in 
\epsffile{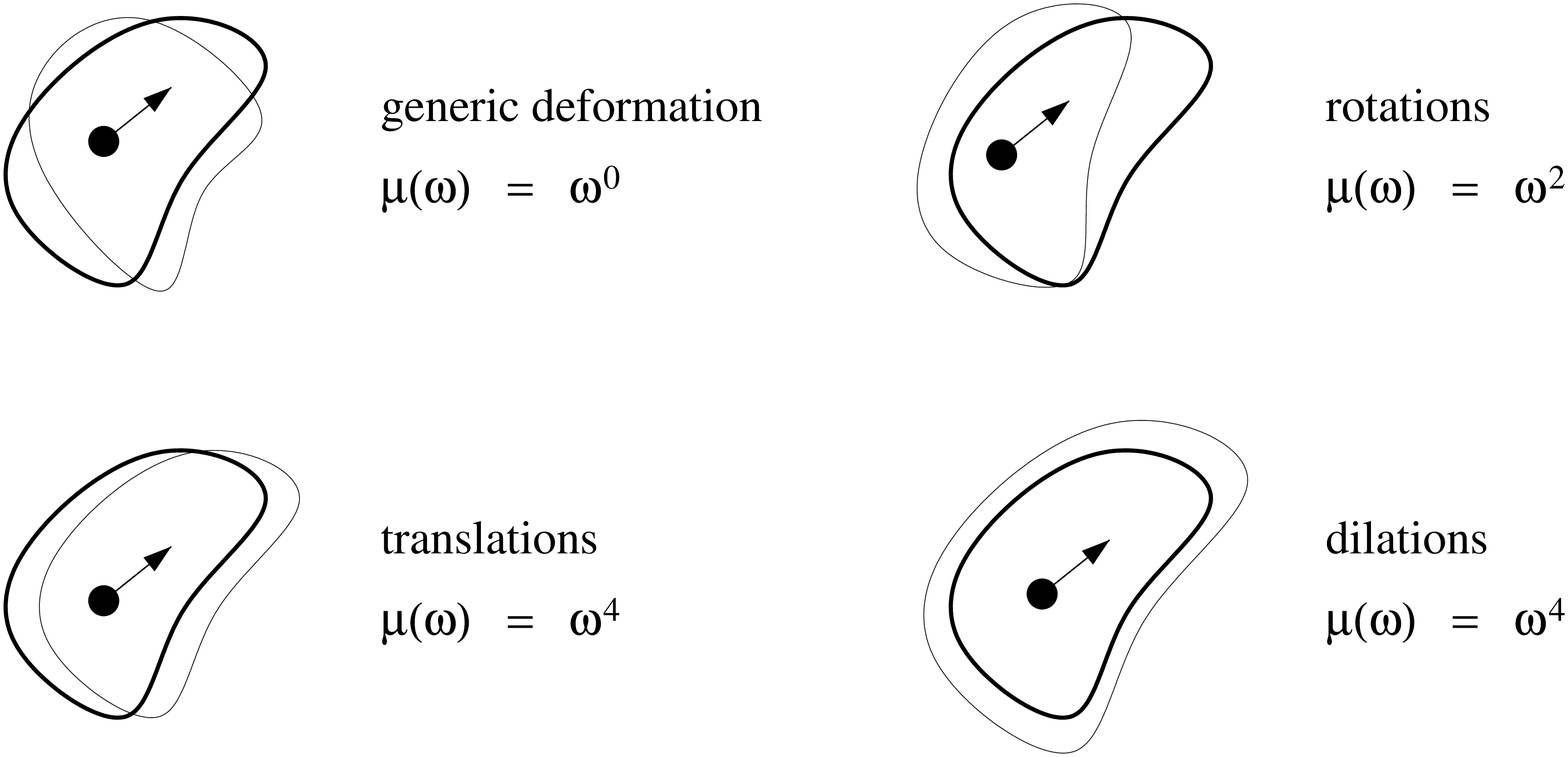}  
\ \\ \ \\ \ \\   
\epsfysize=2.0in 
\epsffile{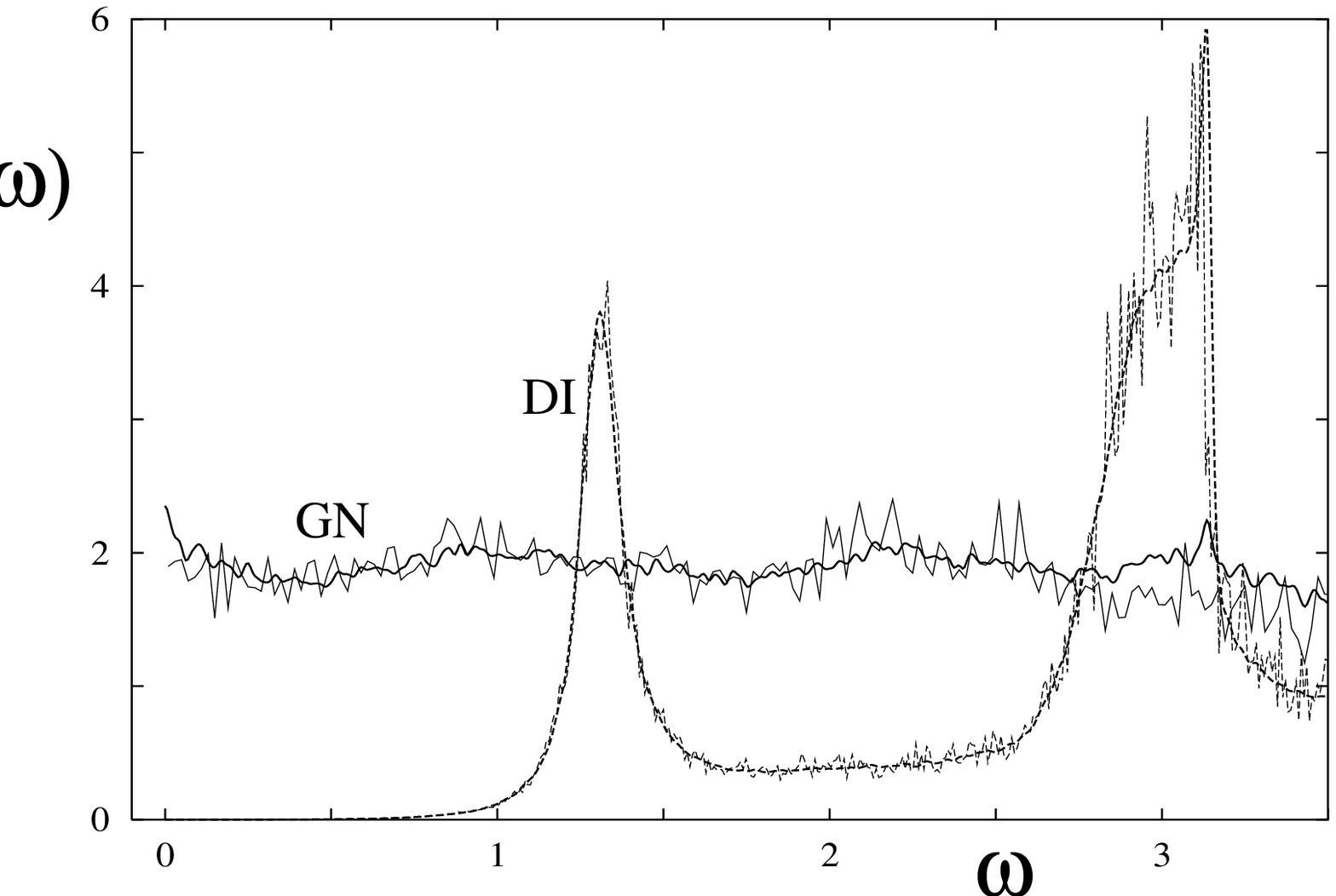}
\end{center}
{\footnotesize {\bf Fig.2.} 
Consider a system of non-interacting particles 
confined inside a chaotic cavity. Assume periodic 
driving $x(t)=A\sin(\omega t)$, where $x$ controls the 
deformation of the boundary. For special type of 
deformations $\mu(\omega)$ vanishes for $\omega \rightarrow 0$. 
A numerical calculation for a stadium shaped cavity 
is displayed on the right. DI is dilation around the 
center, while GN is generic deformation. 
Thick lines are for classical calculation while 
thin lines are for quantum mechanical (linear response) 
calculation. Note the remarkable agreement.} \\ 

A renewed interest in the wall formula is anticipated in 
the field of mesoscopic physics. 
Quantum dots can be regarded as small 2D cavities whose shape 
is controlled by electrical gates. In such case we 
can use our improved version of the wall formula [P6], 
and incorporate other corrections [to be published]
that take the nature of the dynamics into account. \\ 

Driving a quantum dot by time-dependent magnetic field (Fig.3) 
is an obvious option. We already have mentioned 
the ring geometry (Fig.1, right) where the dissipation coefficient
$\mu$ is just the conductance. However, there is nothing 
special about `rings'.  
One may consider a simple two dimensional quantum dot 
driven by a time-dependent {\em homogeneous} magnetic field. 
For the latter geometry it is better not to use the term 
conductance while referring to $\mu$.

\ \\ \ \\
\begin{center}
\leavevmode
\epsfysize=1.6in 
\epsffile{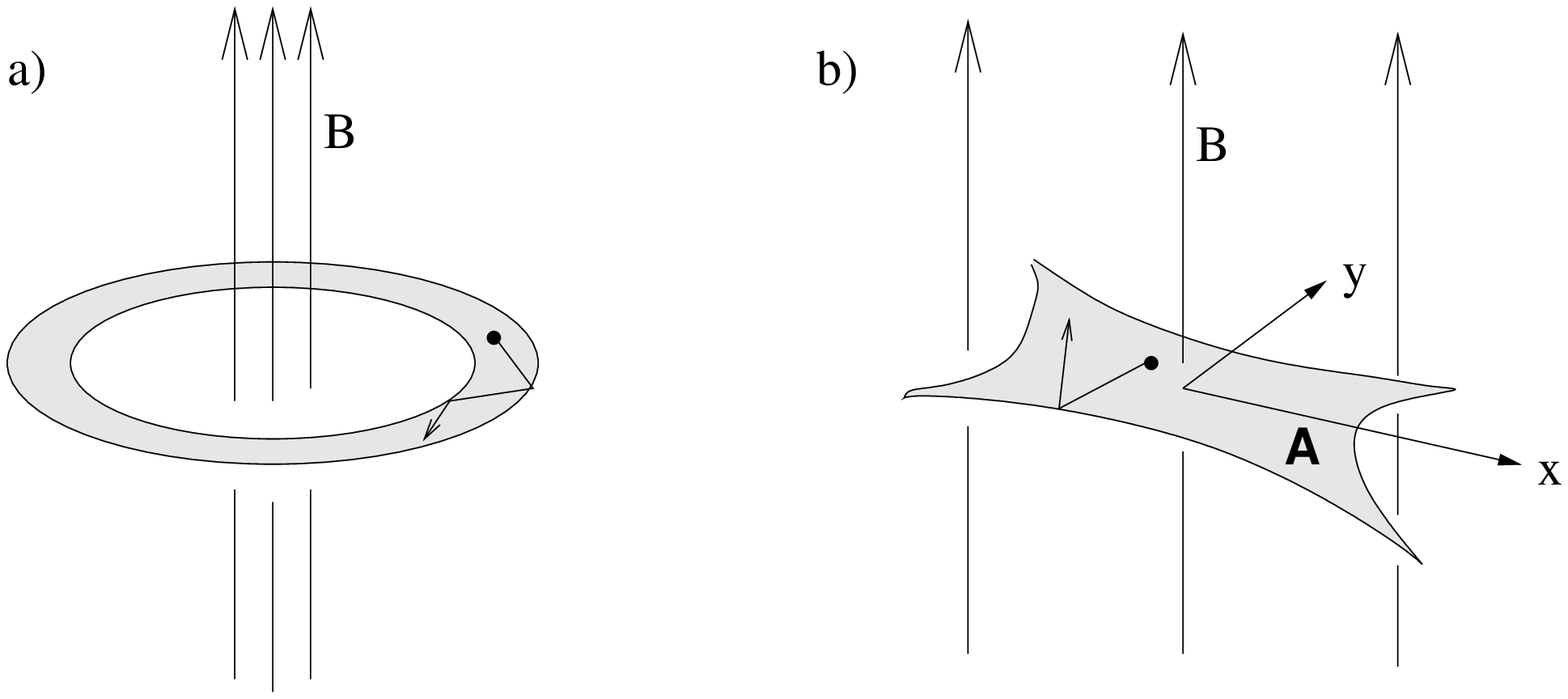} 
\ \\ \ \\ \ \\
\epsfysize=2.7in 
\epsffile{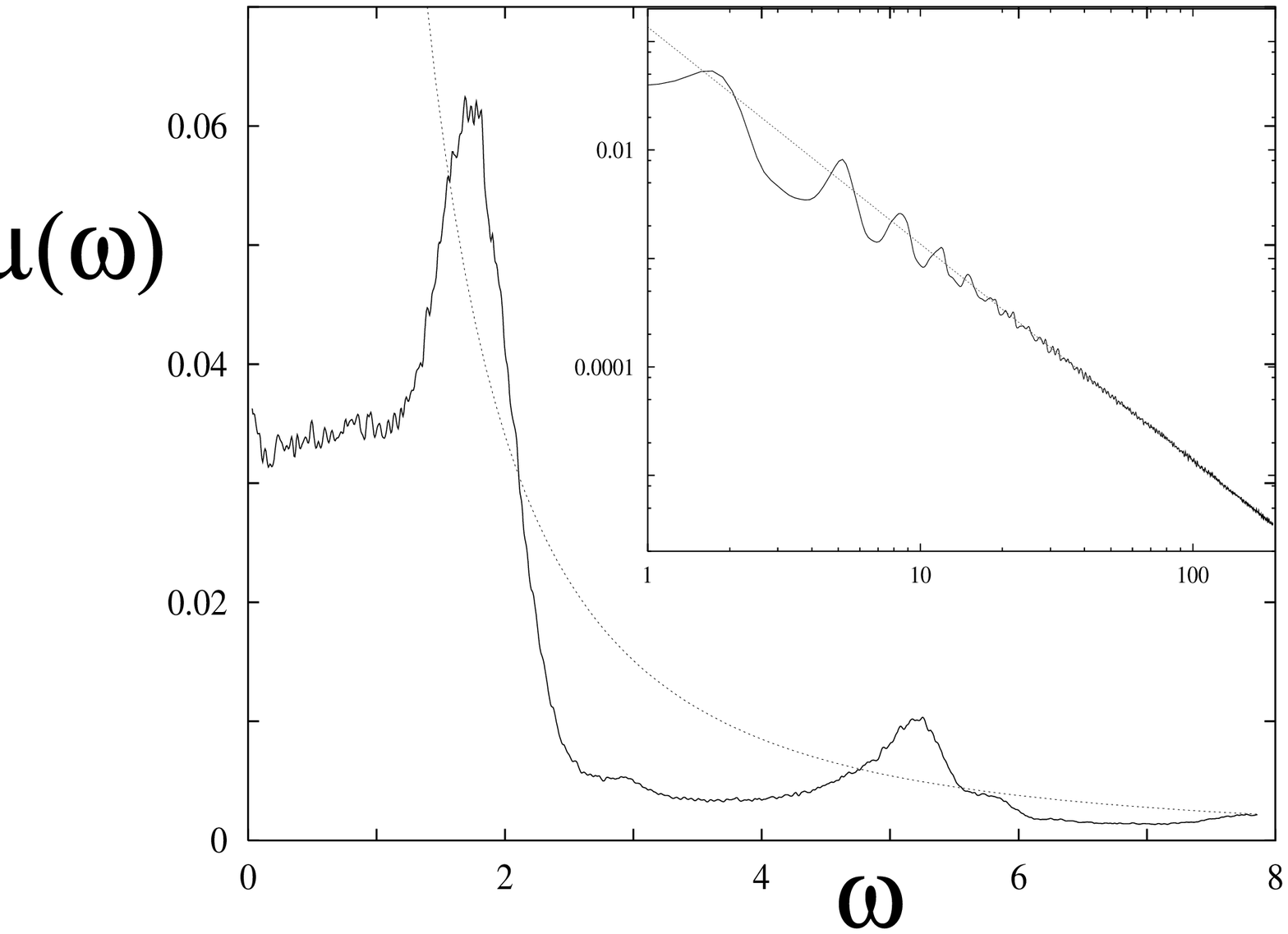}
\end{center}
{\footnotesize {\bf Fig.3.} 
We consider non-interacting electrons driven 
by electro motive force which is induced by a 
time-dependent magnetic field. Rather than 
ring geometry (a) we have a chaotic dot (b). 
The numerical result can be regarded 
as a mesoscopic version of Drude formula. 
The log-log inset on the right demonstrates 
the large frequency $1/\omega^2$ behavior.
} \\ 

It is important to realize that the quantum-mechanical (QM) 
version of linear response theory (LRT) is in remarkable correspondence 
with the classical result (see eg Fig.2). One wonders whether 
QM effects are important. This subject has been addressed 
in [P1-P4]. The main observation is that in the theory 
of {\em quantum} dissipation there are {\em three} distinct 
regimes (Fig.4):

\begin{itemize} 
\item The QM-adiabatic regime.  
\item The linear response (Kubo) regime. 
\item The non-perturbative regime. 
\end{itemize}

\ \\ \ \\ 

\begin{center}
\leavevmode
\hspace*{0in}
\epsfysize=0.5in 
\epsffile{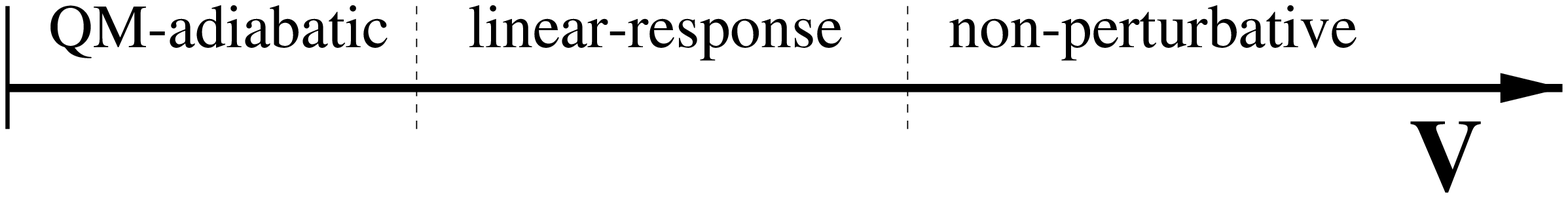} 
\end{center}
{\footnotesize {\bf Fig.4.} 
This diagram illustrates the various $V$ regimes in the 
theory of quantum dissipation for linear driving $x(t)=Vt$.
The more complicated diagram for the case of periodic driving 
is presented in Fig.5 and discussed below. } \\  

\ \\

Past studies of quantum dissipation were 
focused on the QM-adiabatic regime 
(extremely slow driving), and have 
dealt with either the Landau-Zener mechanism~[5,6]
or else with the Debye relaxation absorption mechanism~[7]
for dissipation. The appearance of the QM-adiabatic regime 
is related to the existence of a finite mean level spacing $\Delta$. \\

The surprising message of [P1-P4] is that 
there is a new regime in the theory of quantum dissipation, 
where QM LRT fails. This failure is not related to 
having finite mean level spacing $\Delta$, 
but rather to having finite bandwidth $b\times\Delta$ 
of the perturbation matrix. 
A well known semiclassical relation~[8]
relates the bandwidth to the dropoff frequency of 
the classical $\mu(\omega)$. Namely   
$b\Delta = \hbar \omega_{\tbox{cl}}$. 
In the context of mesoscopic physics the bandwidth 
is known as the Thouless energy.     
Another observation of [P1-P3] is that the 
semiclassical regime is contained inside 
the non-perturbative regime.  \\

The various $(\omega,A)$ regimes for 
periodic driving $x(t)=A\sin(\omega t)$  
are illustrated in Fig.5.  
The QM-adiabatic regime (excluding the narrow stripes 
of resonances) is defined by having {\em vanishing} 
first-order probability to go to other levels.  
In the LRT regime it is assumed that there is 
strong response for $\omega<\omega_{\tbox{cl}}$ 
and vanishingly small response otherwise.  
Quantal non-perturbative response appears provided 
the driving amplitude $A$ is large enough. 
The theory has been tested [P4] in collaboration 
with {\em T. Kottos} for a RMT model.  \\

\newpage

\ \\  
\begin{center}
\leavevmode
\epsfysize=2.7in 
\epsffile{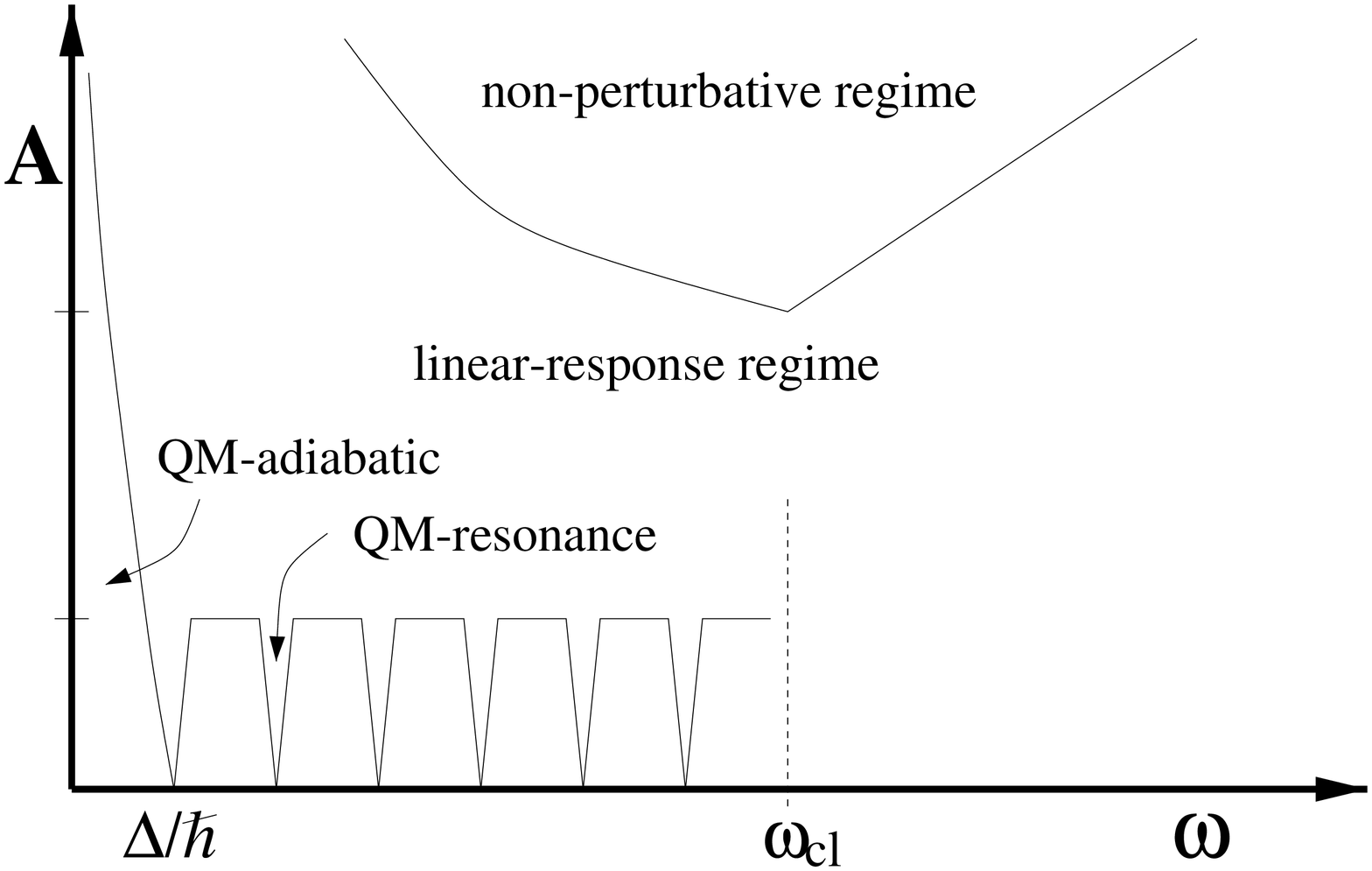} 
\ \\ \ \\ \ \\ \ \\ 
\epsfysize=3.5in 
\epsffile{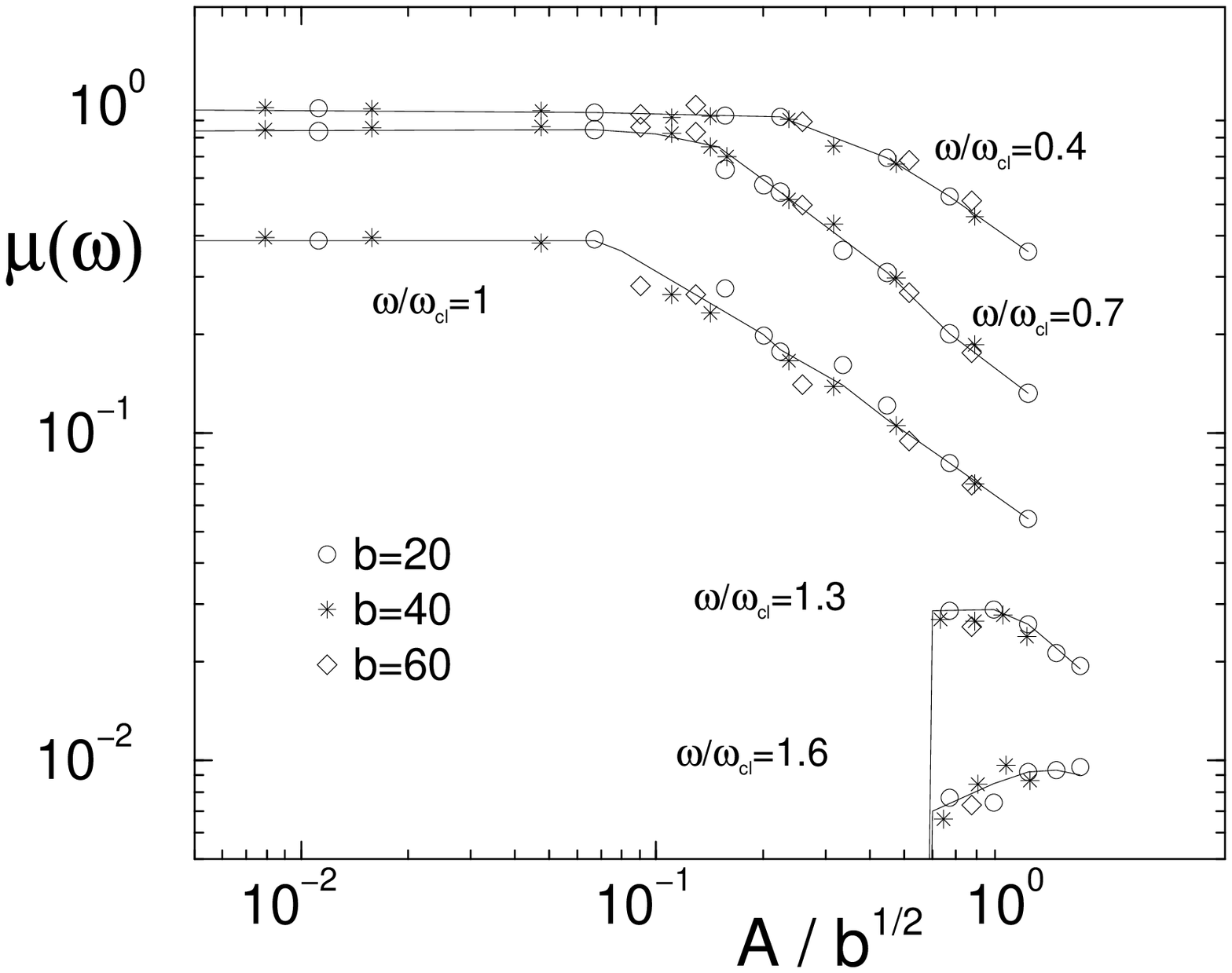}
\end{center}
{\footnotesize {\bf Fig.5.} 
{Upper:} 
Diagram illustrating the various $(\omega,A)$ regimes in the 
theory of quantum dissipation for periodic driving. 
{Lower:}  
The results of RMT simulations (Wigner model). 
The dependence of $\mu(\omega)$ on the driving amplitude $A$ is displayed. 
$\mu=\mbox{const}$ behavior is implied by LRT. 
The observed failure of LRT, and the horizontal scaling with respect 
to the bandwidth $b$, are in accordance with the theoretical 
expectations. } \\ 

\newpage

\ \\ \ \\ \ \\
{\bf References:}

\begin{description}

\item[{\rm [1]}]
J. Blocki et all. {\em Ann. Phys.} {\bf 113}, 330 (1978).

\item[{\rm [2]}]
M. Wilkinson, {\em J. Phys. A {\bf 23}}, 3603 (1990).

\item[{\rm [3]}]
C. Jarzynski, {\em Phys. Rev.} {\bf E 48}, 4340 (1993).

\item[{\rm [4]}]
E. Ott, {\em Phys. Rev. Lett.} {\bf 42}, 1628 (1979).

\item[{\rm [5]}]
Y. Gefen and D.J. Thouless, {\em Phys. Rev. Lett.} {\bf 59}, 1752 (1987).

\item[{\rm [6]}]
M. Wilkinson, {\em J. Phys. A} {\bf 21}, 4021 (1988).

\item[{\rm [7]}]
R. Landauer and M. Buttiker, {\em Phys. Rev. Lett.} {\bf 54}, 2049 (1985).

\item[{\rm [8]}]
M. Feingold and A. Peres, {\em Phys. Rev. A} {\bf 34}, 591 (1986).

\end{description}

\ \\ \ \\
{\bf Publications:}

\begin{description}

\item[{\rm [P1]}] 
D. Cohen, "Quantum Dissipation due to the interaction with 
chaotic degrees-of-freedom and the correspondence principle", 
{\it Phys. Rev. Lett. {\bf 82}}, 4951 (1999).

\item[{\rm [P2]}]         
D. Cohen, "Chaos and Energy Spreading for Time-Dependent Hamiltonians, 
and the various Regimes in the Theory of Quantum Dissipation", 
{\it Annals of Physics {\bf 283}}, 175 (2000). 

\item[{\rm [P3]}]        
D. Cohen, "Chaos, Dissipation and Quantal Brownian Motion", 
Proceedings of the International School of Physics 
`Enrico Fermi' Course CXLIII ``New Directions in Quantum Chaos'', 
Edited by G. Casati, I. Guarneri and U. Smilansky, 
IOS Press, Amsterdam (2000). 

\item[{\rm [P4]}] 
D. Cohen and T. Kottos, "Quantum-mechanical non-perturbative response 
of Driven Chaotic Mesoscopic Systems, cond-mat/0004022, 
{\it Phys. Rev. Lett.} (2000, in press). 

\item[{\rm [P5]}]
A. Barnett, D. Cohen and E.J. Heller, "Deformations and dilations 
of chaotic billiards, rate of dissipations, and quasi-orthogonality 
of the boundary wavefunctions", 
{\it Phys. Rev. Lett. {\bf 85}}, 1412 (2000).
 
\item[{\rm [P6]}]
A. Barnett, D. Cohen and E.J. Heller, 
"Rate of energy absorption for a driven chaotic cavity", 
nlin.CD/0006041, {\it J. Phys. A} (2000, to be published).

\end{description}
\ \\
\ \\ 
\rule{10cm}{.01in}

\noindent
{\footnotesize Preprints and publications are available via \\
{\em http://monsoon.harvard.edu/$\sim$doron}}

\end{document}